\RequirePackage{fix-cm}

\documentclass[smallextended]{svjour3}       
\usepackage{xcolor}
\usepackage{graphicx}
\usepackage{subfigure}
\usepackage{amssymb,amsmath}
\usepackage{doi}
\usepackage{hyperref}
\usepackage[normalem]{ulem}
\hypersetup{colorlinks,linkcolor=black,citecolor=blue,urlcolor=blue}

\makeatletter
\newenvironment{highlightequation}
{\def\tagform@##1{\maketag@@@{(\ignorespaces##1\unskip\@@italiccorr*)}}%
\ignorespaces}
{\def\tagform@##1{\maketag@@@{(\ignorespaces##1\unskip\@@italiccorr)}}%
\ignorespacesafterend}
\makeatother

\DeclareMathOperator{\tr}{tr}

\graphicspath{{/}{Figs/}}

\begin{document}


\title{``Wunderlich, meet Kirchhoff'':\\
A general and unified description 
of elastic ribbons and thin rods}

\author{Marcelo A. Dias\and
        Basile Audoly }
\institute{M. A. Dias \at
	   School of Engineering, Brown University, Providence, RI 
	   02912, USA\\
	   \email{marcelo$\underline{\,\,\,\,}$dias@brown.edu}
	   \and
	   B. Audoly \at
           Sorbonne Universit\'es, UPMC Univ Paris 06, CNRS,
	   UMR 7190 Institut Jean Le Rond d'Alembert, F-75005 Paris, France\\
	   \email{audoly@lmm.jussieu.fr}
}

\date{Received: date / Accepted: date}

\maketitle

\begin{abstract}

The equations for the equilibrium of a thin elastic ribbon are derived
by adapting the classical theory of thin elastic rods.  Previously
established ribbon models are extended to handle geodesic curvature,
natural out-of-plane curvature, and a variable width.  Both the case
of a finite width (Wunderlich's model) and the limit of small width
(Sadowksky's model) are recovered.  The ribbon is assumed to remain
developable as it deforms, and the direction of the generatrices is
used as an internal variable.  Internal constraints expressing
inextensibility are identified.  The equilibrium of the ribbon is
found to be governed by an equation of equilibrium for the internal
variable involving its second-gradient, by the classical Kirchhoff
equations for thin rods, and by specific, thin-rod-like constitutive
laws; this extends the results of Starostin and van der Heijden (2007)
to a general ribbon model.  Our equations are applicable in particular
to ribbons having geodesic curvature, such as an annulus cut out in a
piece of paper.  Other examples of application are discussed.  By
making use of a material frame rather than the Frenet--Serret
frame, the present work unifies the description of thin ribbons
and thin rods.  

\end{abstract}


\section{Introduction}

A ribbon is an elastic body whose dimensions
(typical length $L$, width $w$ and thickness $h$) are all very
different, 
$L\gg w \gg h$.  
While previous work has been focussed on the case of rectangular
ribbons, we consider the general case of ribbons having non-zero
natural curvatures, both in the out-of-plane and the in-plane
directions.  This extension includes ribbon geometries such as those
obtained by cutting a piece of paper along two arbitrary curves.

From a mechanical perspective, elastic ribbons lie halfway between the
1D case of thin rods (for which $w\sim h$), and the 2D case of thin
elastic plates or shells (for which $w\sim L$).  
On one hand, their elastic energy is given by the theory of thin
elastic plates or shells.  On the other hand, ribbons look like 1D
structures (thin rods) when observed from the large scale $L$: this
suggests that they can be described by the classical equations for
thin elastic rods.  This article is concerned with the following
problem of dimensional reduction: starting from a thin, developable
shell model, can one recover the 1D equations of equilibrium
applicable to thin rods?

This work builds up on a few seminal articles.  The dimensional
reduction has already been carried out at the energy level and in the
particular case of rectangular, naturally flat ribbons:
Sadowsky~\cite{Sadowsky30} derived a 1D energy functional for a narrow
ribbon (small $w$), and his work was later generalized by
Wunderlich~\cite{Wunderlich1962} to a finite width $w$.  Their
dimensional reduction was made possible by focussing on developable
configurations of the ribbon, which are preferred energetically in the
thin limit, $h\ll w$.  Developable surfaces are special cases of ruled
surfaces, \emph{i.e.}\ they are spanned by a set of straight lines
called generatrices or rulings: the 1D elastic energy of Wunderlich is
based on a reconstruction of the surface of the ribbon in terms of its
center-line and of the angle between the generatrices and the
center-line tangent.  We use a similar parameterization here and
derive the 1D energy functional for a developable, but not necessarily
rectangular, ribbon.

Next comes the question of minimizing this 1D energy to solve the
equilibrium problem.  Upper bounds for the energy have been obtained by
inserting trial forms of the ribbon into the 1D energy, as was done in
the context
of the elastic M\"obius strip~\cite{Wunderlich1962,%
MahadevanKeller-The-shape-of-a-Mobius-band-1993}.  Finding equilibrium
solutions, however, requires one to derive the equations of
equilibrium by a variational method.  This has been done in a
beautiful article by Starostin and van der
Heijden~\cite{Starostin2007} for naturally flat and rectangular
ribbons.  They found equilibrium equations that bear a striking
resemblance with the Kirchhoff equations governing the equilibrium of
thin rods.  Their result was later extended to helical
ribbons~\cite{Starostin2008}, which is another case where geodesic curvature is
absent.  Here, we want to revisit and extend their work in the
following ways.

First, the derivation of Starostin and van der Heijden, based on the
variational bicomplex formalism, uses a different approach than the
classical theory of thin rods.  The final equations, however, look
similar to the Kirchhoff equations for the equilibrium of thin rods.
In fact, previous work on thin ribbons has developed as a field
largely independent from the vast literature on thin rods.  This is
unfortunate in view of their deep similarities.  Here, we advocate the
viewpoint that a ribbon is just a special kind of a thin rod, having an
internal parameter and being subjected to kinematical constraints ---
this is quite similar to the way the incompressibility constraint is
handled in 3D elasticity.  These specificities can be incorporated
naturally into the classical theory of thin rods, as we show.  Doing
this allows one to recycle much of the existing knowledge on thin
rods.  In particular, the equations of equilibrium for ribbons are
derived in close analogy with those for rods, and in a straightforward
way.

Second, we make use of directors, as in the classical theory of rods.
By contrast, Wunderlich has introduced a parameterization of the
mid-surface of the ribbon based on the Frenet--Serret frame
associated with the center-line.  Wunderlich's energy, in particular,
is defined in terms of the Frenet--Serret notions of torsion and
curvature.  This parameterization has a drawback: it is specific to
the case where the center-line is a geodesic, as we show.  By working
instead with directors, we can
naturally extend Wunderlich's model to ribbons that have geodesic
curvature, \emph{i.e.} to ribbons curved in their own plane such as an
annulus cut out from a piece of paper.

In the present work we make use of several ideas introduced in a
recent article~\cite{Dias2013}, where we have shown that the buckling
of a curved strip cut out from a piece of paper and folded along its
central circle~\cite{Dias2012a} can be analyzed using the language of
thin rods.  This was done by identifying the relevant geometrical
constraints and constitutive laws.  Here, we do not consider any fold
but allow for more general geometries (non-uniform width and geodesic
curvature).

This article is organized as follows.  In section~\ref{sec:geometry},
we extend the parameterization of developable surfaces introduced by
Wunderlich: making use of the frame of directors, we account for the
geodesic curvature of the center-line and a variable width.  In
section~\ref{sec:energy}, Wunderlich's energy functional is extended.
In section~\ref{sec:equilibrium}, the equilibrium equations of a
general ribbon are derived by a variational method adapted from the
theory of thin rods.  In section~\ref{sec:specialcases}, we recover
known ribbon models in the special case of a geodesic center-line
($\kappa_{\mathrm{g}} = 0$) and constant width $w$.  In
section~\ref{sec:illustrations}, we present some equilibrium problems
for ribbons having geodesic curvature as possible illustrations of our
theory.

\section{Geometry of a developable ribbon}
\label{sec:geometry}

\subsection{Developable transformation from reference to current configuration}
\label{ssec:devtransf}

\begin{figure}
    \centering
    \includegraphics[width=4.5in]{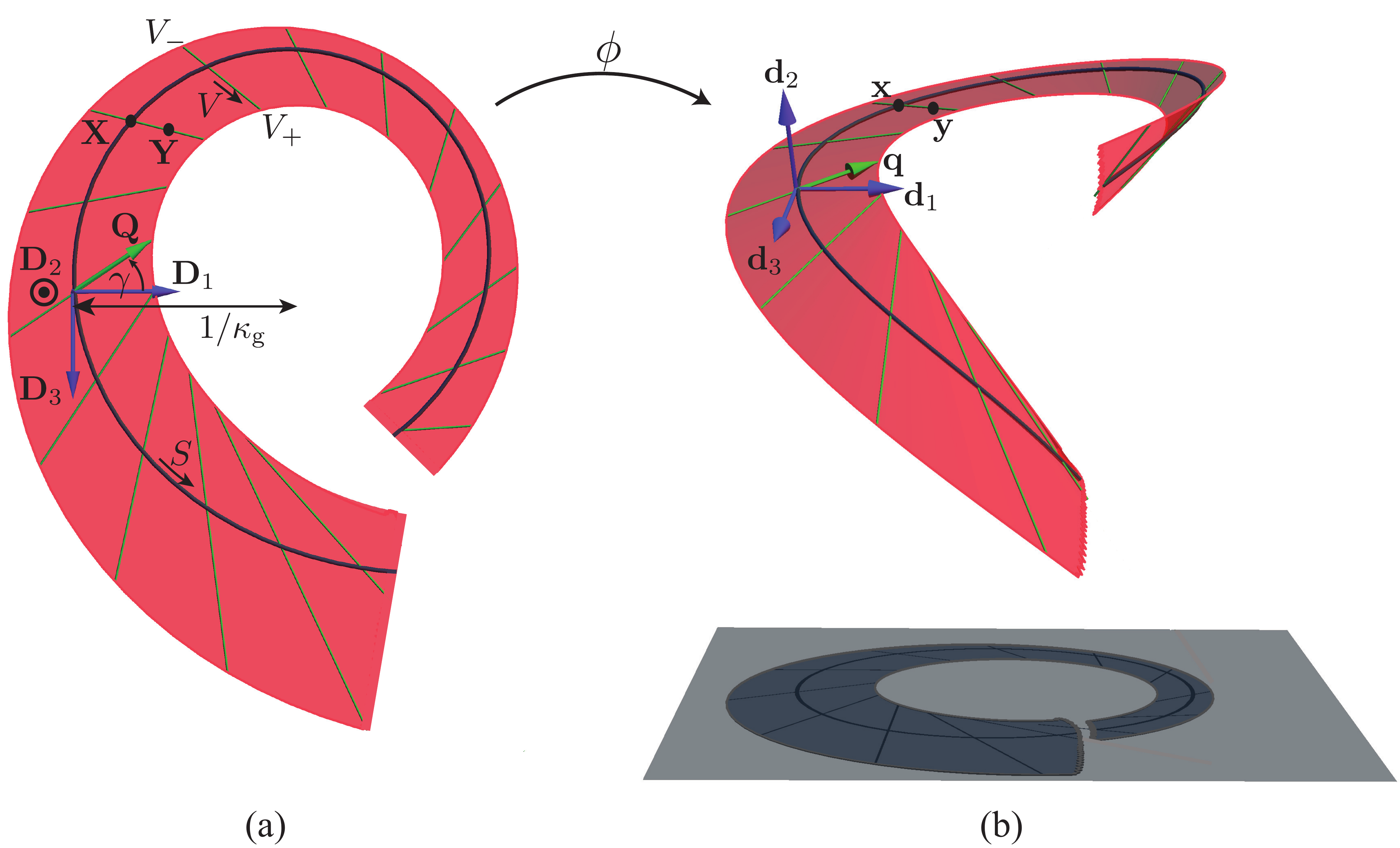}
    \caption{\label{fig:Fig1} Geometry of a developable ribbon (a) in
    the planar, undeformed configuration, and (b) in the actual
    configuration.  The direction of the generatrices is measured by
    the parameter $\eta$ in the model, which is the tangent of the
    angle $\gamma$ between the director $\mathbf{d}_{1}$ and the
    generatrix direction $\mathbf{q}$ (the angle $\gamma$ shown in the
    figure is negative, and $\eta < 0$ here).}
\end{figure}

As we consider developable ribbons, we can assume that the reference
configuration is planar\footnote{For a closed developable ribbon,
there may not exist any \emph{global} planar configurations --- see
the example in section~\ref{ssec:overcurvedAnnular}.  In that case, we
introduce an arbitrary cut in the planar configuration of reference.}.
This planar reference configuration is not necessarily stress-free (we
shall address the case of naturally curved ribbons).  In the reference
configuration, a material line $\mathbf{X}(S)$, called the
\emph{center-line}, is traced out on the ribbon, see
figure~\ref{fig:Fig1}a.
Here, $S$ is the arc-length along the center-line as measured in
reference configuration, $|\mathbf{X}'(S) = 1|$.  Primes denote
derivation with respect to arc-length, and boldface characters denote
vectors.  The surface of the ribbon is oriented by prescribing a
constant unit vector $\mathbf{D}_{2}$, perpendicular to the plane of
the ribbon.  Let $\mathbf{D}_{3}(S) = \mathbf{X}'(S)$ be the unit
tangent to the center-line.  The vectors $\mathbf{D}_{1}(S) =
\mathbf{D}_{2} \times \mathbf{D}_{3}(S)$, $\mathbf{D}_{2}$ and
$\mathbf{D}_{3}(S) $ then form an orthonormal frame.

A deformed, developable configuration of the ribbon is specified by
the functions
\begin{equation}
    \big(
    \mathbf{x}(S), \mathbf{d}_{1}(S), \mathbf{d}_{2}(S),
    \mathbf{d}_{3}(S), \eta(S)
    \big),
    \label{eq:unknowns}
\end{equation}
which are subjected to geometrical constraints derived later.  Here
$\mathbf{x}(S)$ is the deformed center-line, $\mathbf{d}_{i}(S)$ (for
$i=1,2,3$) define the \emph{frame of directors} (also called the
\emph{material frame}) and $\eta(S)$, defined below, encodes the
definition of the generatrices.

The third director is chosen to be the tangent to the deformed
center-line,
\begin{highlightequation}
\begin{equation}
    \mathbf{x}'(S) = \mathbf{d}_{3}(S),
    \label{eq:adaptation}
\end{equation}
\end{highlightequation}
and the second director $\mathbf{d}_{2}(S)$ is defined to be normal to
the ribbon at $\mathbf{x}(S)$ as in the reference configuration, see
figure~\ref{fig:Fig1}b.  The directors are defined to be orthonormal,
\begin{highlightequation}
\begin{equation}
    \mathbf{d}_{i}(S)\cdot \mathbf{d}_{j}(S) = \delta_{ij}
    \textrm{,}
    \label{eq:orthonormal}
\end{equation}
\end{highlightequation}
where $\delta_{ij}$ stands for Kronecker's symbol.  This implies that 
$|\mathbf{x}'| = |\mathbf{d}_{3}| = 1= |\mathbf{X}'(S)|$.
By construction, the directors $\mathbf{d}_{i}$ are \emph{material}\
vectors: contrary to the Frenet--Serret frame associated with the
center-line, they follow the rotation of the ribbon.

As the ribbon is inextensible, it remains developable by Gauss'
\emph{Theorema egregium}.  Smooth, developable surfaces are 
ruled~\cite{Spivak-A-comprehensive-introduction-vol3-1999}:
there exists a one-parameter family of straight lines, called
\emph{generatrices}, that sweeps out over the entire surface.  As in
previous work~\cite{Wunderlich1962,Starostin2007}, we define
$\eta(S)$ to be the tangent of the angle $\gamma$ between
$\mathbf{d}_{1}$ and the generatrix.  Then, the vector
\begin{equation}
    \label{eq:EmbedRibbon2}
    \mathbf{q}(\eta,S) = \eta(S)\,\mathbf{d}_3(S)+\mathbf{d}_1(S)
\end{equation}
spans the generatrix\footnote{The vector $\mathbf{q}$ depends both on
the unknown \emph{function} $\eta(\cdot)$ and on the arc-length
parameter $S$; hence the arguments shown in the left-hand side of
equation~(\ref{eq:EmbedRibbon2}).}.  Therefore, the transformation
from the reference to the deformed configuration can be expressed as
the mapping $\phi$:
\begin{equation}
    \phi:\quad
    \mathbf{Y} = \mathbf{X}(S) + V\,\mathbf{Q}(\eta,S) \;\;\mapsto\;\;
    \mathbf{y} = \mathbf{x}(S) + V\,\mathbf{q}(\eta,S)
    \textrm{.}
    \label{eq:EmbedRibbon}
\end{equation}
Here, $V$ is a coordinate along the generatrix, $\mathbf{Y}$ and
$\mathbf{y}$ denote a current point along the ribbon in reference and
actual configurations, respectively.  The vector $\mathbf{Q}$ is
defined as $\mathbf{Q}(\eta,S) = \eta(S)\,\mathbf{D}_{3}(S) +
\mathbf{D}_{1}(S)$: it defines the direction of the generatrix brought
back in the reference configuration.

We use the longitudinal and transverse coordinates $(S,V)$ to
parameterize the ribbon's surface.  $S$ varies in the interval $0\leq
S\leq L$, where $L$ is the curvilinear length of the center-line.  The
transverse coordinate $V$ varies in a domain $V_{-}(\eta,S) \leq V
\leq V_{+}(\eta,S)$.  The endpoints $V_{\pm}(\eta,S)$ of the interval
are such that the points $\mathbf{Y}_{\pm}(S) = \mathbf{X}(S) +
V_{\pm}(\eta,S)\,\mathbf{Q}(\eta,S)$ lie on the edges of the ribbon.
The functions $V_{\pm}(\eta,S)$ capture the relative position of the
edges and of the center-line, and are called the \emph{edge
functions}.  Explicit expressions are derived in
section~\ref{ssec:VpmFunctions} for some ribbon geometries.


From equation~(\ref{eq:orthonormal}), the directors define an
orthonormal frame for any value of the arc-length parameter $S$.
Therefore, there exists a vector $\boldsymbol{\omega}(S)$ called the
Darboux vector or the rotation gradient, such that
\begin{equation}
    \mathbf{d}_{i}'(S) = \boldsymbol{\omega}(S) \times 
    \mathbf{d}_{i}(S)
    \label{eq:Frame1}
\end{equation}
for $i=1,2,3$.  The operation $\times$ denotes the cross product in the
Euclidean space.  The components of the rotation gradient in the basis
of directors, $\omega_{i}(S) = \boldsymbol{\omega}(S)\cdot
\mathbf{d}_{i}(S)$ measure the amount of bending ($i=1,2$) and
twisting ($i=3$) of the center-line. An explicit expression is
\begin{highlightequation}
\begin{equation}
    \label{eq:Frame1-bis}
    \omega_{i}(S) = \frac{1}{2}\,\sum_{j=1}^3\sum_{k=1}^3\epsilon_{ijk}\,
    \mathbf{d}_{j}'(S)\cdot \mathbf{d}_{k}(S)    
    \textrm{,}
\end{equation}
\end{highlightequation}
where $\epsilon_{ijk}$ represents the permutation symbol:
$\epsilon_{ijk} = 1$ when $(i,j,k)$ is an even permutation of the
indices, $\epsilon_{ijk} = -1$ when it is an odd permutation, and
$\epsilon_{ijk} = 0$ otherwise.  In the language of the geometry of
surfaces, the directors frame
$(\mathbf{d}_{1},\mathbf{d}_{2},\mathbf{d}_{3})$ is called the Darboux
frame associated with the center-line curve, and the strains
$\omega_{1}$, $\omega_{2}$, and $\omega_{3}$ are respectively the
normal curvature, the geodesic curvature, and the geodesic torsion.
Note we use strain measures, $\omega_{i}$, that are based on the frame
of directors (a material frame), while previous work used the
Frenet--Serret frame associated with the center-line.  Working with a
frame of directors offers many advantages: it extends naturally to the
case of non-geodesic center-lines, allows one to use the same language
as in the theory of rods, and to remove the artificial singularities
displayed by the Frenet--Serret frame near inflection points or
straight segments.


\subsection{Edge functions}
\label{ssec:VpmFunctions}

The edge functions $V_{\pm}(\eta,S)$ encode the relative positions of the edges of the ribbon with respect to the center-line.  Expressions for $V_{\pm}$ are derived below for the cases of a rectangular and an annular ribbon.

The case of a rectangular ribbon
is quite simple. We use the central axis of the ribbon
as the center-line.  In reference configuration, the equation of the
edges is $(\mathbf{Y} - \mathbf{X})\cdot \mathbf{D}_{1} = \pm w/2$,
where $w$ is the width of the ribbon.  Inserting the parameterization
of $\mathbf{Y}$ from equation~(\ref{eq:EmbedRibbon}), this yields
\begin{equation}
    V_{\pm}(\eta,S) = \pm\frac{w}{2}\quad
    \textrm{(rectangular ribbon)}
    \textrm{,}
    \label{eq:rectangular-ribbon}
\end{equation}
where the $\pm$ labels the edges.

The case of an annular ribbon is studied next.  We use an arc of a
circle as the center-line, having its radius given by the inverse of
the geodesic curvature, $\kappa_{\mathrm{g}}^{-1}$, and a constant
width $w$ of the ribbon.  In reference configuration, the center
$\mathbf{C}$ of the circular center-line is $\mathbf{C} =
\mathbf{X}(S)+\kappa_{\mathrm{g}}^{-1}\,\mathbf{D}_{1}(S)$.
Therefore, the equation for the edges is $(\mathbf{Y}(S,V) -
\mathbf{C})^2 = \left(\kappa_{\mathrm{g}}^{-1} \mp \frac{w}{2}
\right)^2$.  Using equation~(\ref{eq:EmbedRibbon}) and the definition
of $\mathbf{Q}$, this shows that the edge functions $V_{\pm}$ are the
roots $V$ of the second-order polynomial
\begin{equation}
    \left(V-\frac{1}{\kappa_{\mathrm{g}}}\right)^2 + (\eta\,V)^2 = 
    \left(\frac{1}{\kappa_{\mathrm{g}}} \mp \frac{w}{2} 
    \right)^2
    \textrm{.}
    \nonumber
\end{equation}
Solving for $V$, we find
\begin{equation}
    \label{eq:Bounds}
    V_{\pm}(\eta,S) = \frac{1}{\kappa_{\mathrm{g}}}\,
    \frac{1-\sqrt{1\mp(1+\eta^2)\,w\,\kappa_{\mathrm{g}}\,
    \left(1\mp\frac{w\,\kappa_{\mathrm{g}}}{4}\right)}}{1+\eta^2}
    \quad
    \textrm{(annular ribbon)}
    \textrm{.}
\end{equation}
Letting the curvature of the annulus go to zero,
$\kappa_{\mathrm{g}}\to 0$, we recover $V_{\pm}\to \pm w/2$ which is
consistent with equation~(\ref{eq:rectangular-ribbon}).

In the general case of a ribbon having a variable width, or when the
center-line has non-constant curvature $\kappa_{\mathrm{g}}$,
$V_{\pm}$ may be available through an implicit equation and not
necessarily in closed form.

\subsection{Constraints expressing developability}
\label{ssec:constraints}

The condition of inextensibility of the ribbon imposes some
kinematical constraints on the unknowns listed in (\ref{eq:unknowns}),
and on the curvature and twisting strains $\omega_{i}$ calculated by
equations~(\ref{eq:Frame1}--\ref{eq:Frame1-bis}).  These constraints
are derived as follows.

The center-line is a curve drawn on the surface of the ribbon.  Its
geodesic curvature is defined by $\kappa_{\mathrm{g}} =
\mathbf{x}''\cdot \mathbf{d}_{1} = \mathbf{d}_{3}'\cdot \mathbf{d}_{1}
= \omega_{2}$.  It is a classical result of the differential geometry of
surfaces~\cite{Spivak-A-comprehensive-introduction-vol3-1999}
that the geodesic curvature is conserved upon isometric
deformations of a surface.  Therefore, $\kappa_{\mathrm{g}}$ is
prescribed by the reference configuration: $\kappa_{\mathrm{g}}(S) =
\mathbf{D}_{3}'(S)\cdot \mathbf{D}_{1}(S)$.  We write this geodesic
constraint as
\begin{highlightequation}
    \begin{subequations}
    \label{eq:geodesicConstraint}
    \begin{equation}
        \mathcal{C}_{\mathrm{g}}(\omega_{2},S) = 0
	\label{eq:geodesicConstraint-cancels}
    \end{equation}
    where
    \begin{equation}
	\mathcal{C}_{\mathrm{g}}(\omega_{2},S) = 
	\kappa_{\mathrm{g}}(S) - \omega_{2} 
	\textrm{.}
	\label{eq:geodesicConstraint-def}
    \end{equation}
\end{subequations}

Note that $\omega_{2} = 0$ when $\kappa_{\mathrm{g}} = 0$, \emph{i.e.}
when the center-line is a geodesic.  In that case, the derivative of
the tangent is $\mathbf{d}_{3}' = \boldsymbol{\omega}\times
\mathbf{d}_{3} = \omega_{1}\,(-\mathbf{d}_{2})$ and
$(-\mathbf{d}_{2})'\cdot \mathbf{d}_{1} = \omega_{3}$.  Here, we
recognize the definition of the Frenet--Serret frame
$(\mathbf{d}_{3},-\mathbf{d}_{2},\mathbf{d}_{1})$ associated with the
center-line: the Frenet--Serret curvature and torsion are
$\omega_{1}$ and $\omega_{3}$, respectively.  Therefore, in the
particular case when the center-line is a geodesic, our directors
coincide with the Frenet--Serret frame.  It
is much more convenient to work with the directors in general.

The second constraint expresses the developability of the ruled
surface spanned by the generatrices.  It is found in classical
textbooks of
differential geometry~\cite{Spivak-A-comprehensive-introduction-vol3-1999}, and is rederived in Appendix~\ref{SIA}:
\begin{subequations}
    \label{eq:developabilityConstraint}
    \begin{equation}
        \mathcal{C}_{\mathrm{d}}(\omega_{1},\omega_{3},\eta) = 0
        \label{eq:developabilityConstraint-cancels}
    \end{equation}
    where
    \begin{equation}
        \mathcal{C}_{\mathrm{d}}(\omega_{1},\omega_{3},\eta) =
    \eta\,\omega_{1}  - \omega_{3}
    \textrm{.}
        \label{eq:developabilityConstraint-def}
    \end{equation}
\end{subequations}
\end{highlightequation}

\subsection{Area element}

To integrate the elastic energy along the surface of the 
ribbon, we will need the expression for the area element $\mathrm{d}a$. 
It is calculated in Appendix~\ref{SIA} from the Jacobian of the 
transformation $\phi$:
\begin{equation}
	\label{eq:Element}
	\mathrm{d}a=\left|\partial_S\mathbf{y}\times\partial_V\mathbf{y}
	\right|\,\mathrm{d}S\,\mathrm{d}V=
	\left(1-\frac{V}{V_{\mathrm{c}}(\eta,S)}\right)\mathrm{d}S\,\mathrm{d}V
	\textrm{,}
\end{equation} 
where the auxiliary quantity $V_{\mathrm{c}}$ is defined as
\begin{highlightequation}
\begin{equation}
    \label{eq:Vcritical}
    V_\mathrm{c}(\eta,\eta',S) = 
    \frac{1}{(1+\eta^2)\,\kappa_{\mathrm{g}}(S)-\eta'}.
\end{equation}
\end{highlightequation}
Since $\mathrm{d}a = 0$ at $V=V_{\mathrm{c}}$, the transformation
$\phi$ is singular there.  The quantity $V_{\mathrm{c}}$ can be
interpreted as the value of the transverse coordinate $V$ where the
generatrix intersects neighboring generatrices, \emph{i.e.}\
intersects its own caustic, called the \emph{striction
curve}~\cite{Spivak-A-comprehensive-introduction-vol3-1999}.  We assume that the striction curve stays outside of the physical domain, so that the curvature tensor is
nowhere singular on the ribbon: either $V_{\mathrm{c}} < V_{-} \leq V 
\leq V_{+}$ or $V_{\mathrm{c}} > V_{+} \geq V \geq V_{-}$.  This 
implies that
\begin{subequations}
\begin{equation}
    \frac{1 - V_{+}/V_{\mathrm{c}}}{1 - V_{-}/V_{\mathrm{c}}} > 0
    \label{eq:logArgumentPositive1}
 \end{equation}
    and
 \begin{equation}
    \left|
    \frac{V}{V_{\mathrm{c}}}
    \right| < 1
    \label{eq:logArgumentPositive2}
\end{equation}
\end{subequations}

\subsection{Curvature tensor of the deformed ribbon}
\label{ssec:curvatureTensor}

In order to write the elastic energy of the ribbon, we need the
curvature tensor $\mathbf{K}$ at the arbitrary point $(S,V)$ of the
ribbon surface.  This geometrical calculation, at the heart of
Wunderlich's energy, is given in Appendix~\ref{SIA}.  The dependence of
the curvature tensor, $\mathbf{K}(\eta,\omega_{1},S,V)$, on the
transverse coordinate is imposed by the developability condition to be
\begin{equation}
    \label{eq:Curvature2FormMaterial}
    \mathbf{K}(\eta,\eta',\omega_{1},S,V) = 
    \frac{\mathbf{K}_{0}(\eta,\omega_{1})}{1-\frac{V}{V_{\mathrm{c}}(\eta,\eta',S)}}
    \textrm{,}
\end{equation}
where $\mathbf{K}_{0}(\eta,\omega_{1}) = \mathbf{K}(\eta,\eta',\omega_{1},S,V=0)$ denotes the
curvature tensor evaluated along the center-line, defined by
\begin{highlightequation}
\begin{equation}
    \label{eq:Curvature2FormMaterial0}
    \mathbf{K}_{0}(\eta,\omega_{1})=
    -\omega_1\,
   \left(\mathbf{d}_3
    \otimes\mathbf{d}_3
    -\eta\,
    (
    \mathbf{d}_3\otimes\mathbf{d}_1
    +
    \mathbf{d}_1\otimes\mathbf{d}_3
    )
    +\eta^2\mathbf{d}_1\otimes\mathbf{d}_1\right)
     \textrm{.}
\end{equation}
\end{highlightequation}
Here, the curvature tensor is expressed by its components in the
orthonormal basis $(\mathbf{d}_{3}(S),\mathbf{d}_{1}(S))$ spanning the
plane tangent to the ribbon at $(S,V)$, as implied by the subscript
notation.  Inserting equation~(\ref{eq:Curvature2FormMaterial0}) into
equation~(\ref{eq:Curvature2FormMaterial}) yields $\det\mathbf{K} =
\det\mathbf{K}_{0 } = 0$ which is consistent with Gauss'
\emph{theorema egregium} (conservation of Gauss curvature by isometric
deformations of a surface).

\section{Elastic energy}
\label{sec:energy}

Being modeled as an inextensible plate, the deformed ribbon is
isometrically mapped to its planar underformed configuration.  The
constraint of isometric mapping is enforced by the
equations~(\ref{eq:geodesicConstraint}--\ref{eq:developabilityConstraint}).
As a result, the stretching energy cancels.  The only contribution to
the elastic energy of the ribbon is the bending energy ${E}$.  For a
plate made up of a homogeneous isotropic Hookean solid,
\begin{multline}
    \nonumber
    {E} 
    =\frac{D}{2}\,
    \iint\Big[(1-\nu)\,\tr\left((\mathbf{K}-\mathbf{K}_\mathrm{r})^2\right)
    +\nu\,\tr^2(\mathbf{K}-\mathbf{K}_\mathrm{r})
    \cdots\\
    {}-(1-\nu)\,\tr(\mathbf{K}_\mathrm{r}^2)-\nu\,\tr^2\mathbf{K}_\mathrm{r}\Big]\,\mathrm{d}a
\end{multline}
where $D=Y\,h^3/(12(1-\nu^2)))$ is the bending modulus of the plate,
$\nu$ the Poisson's ratio, $Y$ the Young's modulus, $h$ the thickness
of the ribbon.  This expression is based on the classical formula for
plates~\cite{Love}, and has been modified to take natural (or
reference) curvature into account through the natural curvature tensor
$\mathbf{K}_{\mathrm{r}}$.  The last two constant terms appearing in
the second line are included to make the energy density zero when the
ribbon is flat ($\mathbf{K} = \mathbf{0}$), which is a convenient
convention.

The energy can be simplified by expanding the squared matrices, using
the identity
$\tr^2\mathbf{K}=\tr\left(\mathbf{K}^2\right)+2\,\det\mathbf{K}$ valid
for $2\times 2$ matrices, and the developability condition $\det
\mathbf{K} = 0$ that follows from
equation~\eqref{eq:Curvature2FormMaterial}:
\begin{equation}
    {E} =\frac{D}{2}\iint\left(\tr\left(\mathbf{K}^2\right)
    -2\,
    \mathbf{Q}_{\mathrm{r}}
    :\mathbf{K}\right)\,\mathrm{d}a
    \textrm{,}
    \label{eq:2Dplateenergy}
\end{equation}
where the symmetric tensor $\mathbf{Q}_{\mathrm{r}}$ capturing the 
effect of the reference curvature is defined by
\begin{equation}
    \label{eq:MomentQ}
    \mathbf{Q}_\mathrm{r}=\mathbf{K}_\mathrm{r}+\nu\,\mathrm{cof}\,\mathbf{K}_\mathrm{r}
    \textrm{,}
\end{equation}
and $\mathrm{cof}\,\mathbf{K}_\mathrm{r}$ denotes the cofactor matrix:
\begin{equation}
    \mathrm{cof}\,\mathbf{K}_\mathrm{r}
    =
   \left(K^\mathrm{r}_{11} \mathbf{d}_1\otimes\mathbf{d}_1
   -K^\mathrm{r}_{13}\,
   (
   \mathbf{d}_1\otimes\mathbf{d}_3
   +
   \mathbf{d}_1\otimes\mathbf{d}_3
   )
   +K^\mathrm{r}_{33}\mathbf{d}_3
    \otimes\mathbf{d}_3\right)
    \textrm{.}
    \label{eq:cofactors}
\end{equation}
The symbols $K^{\mathrm{r}}_{ij}$ denote the components of the natural
curvature $\mathbf{K}_{\mathrm{r}}$ in the tangent basis
$(\mathbf{d}_{3},\mathbf{d}_{1})$, with $i,j=3,1$.

Inserting the expression of the area element $\mathrm{d}a$ from
equation~\eqref{eq:Element} and the known dependence of $\mathbf{K}$
on the transverse coordinate from
equation~(\ref{eq:Curvature2FormMaterial}) into equation~(\ref{eq:2Dplateenergy}) gives
\begin{equation}
    \nonumber
    {E} 
    =\frac{D}{2}\,\int_0^L\left[
    \tr\left({\mathbf{K}_{0}}^2\right)\int_{V_-}^{V_+}\frac{\mathrm{d}V}{1-V/V_{\mathrm{c}}}
    -2\,\mathbf{Q}_\mathrm{r}:\mathbf{K}_{0}\,\int_{V_-}^{V_+}\mathrm{d}V\right]\;\mathrm{d}S.
\end{equation}
Here, by factoring the natural curvature tensor
$\mathbf{Q}_{\mathrm{r}}$ out of the integral along the generatrices, we have
assumed that $\mathbf{Q}_{\mathrm{r}}$ varies on the typical
length-scale $L$ (length of the ribbon) and thus can be
considered constant on the much smaller length-scale $w$ (width of the
ribbon).  Our model thus handles non-uniform geometries
$\mathbf{Q}_{\mathrm{r}}(S)$, even though we omit the argument $S$ in
$\mathbf{Q}_{\mathrm{r}}$ for the sake of legibility.

Integrating along $V$, we obtain the one-dimensional energy functional
\begin{highlightequation}
\begin{multline}
    \label{eq:Energy4}
    {E}(\eta,\eta',\omega_{1}) =\frac{D}{2}\,\int_0^L\Bigg[
    \left(-V_{\mathrm{c}}\,\ln\frac{1-V_+/V_{\mathrm{c}}}{1-V_-/V_{\mathrm{c}}}\right)\,
    \tr\left({\mathbf{K}_{0}}^2\right)\cdots
    \\
    {}-2\,(V_+-V_-)\,\mathbf{Q}_\mathrm{r}:\mathbf{K}_{0}\Bigg]\;\mathrm{d}S.
\end{multline}
\end{highlightequation}
This functional extends Wunderlich's result to the case of a ribbon
with natural out-of-plane curvature (through the term depending on
$\mathbf{Q}_{\mathrm{r}}$), a variable width (through the functions
$V_{\pm}$), and geodesic curvature (through the dependence of
$V_{\pm}$ and $V_{\mathrm{c}}$ on $\kappa_{\mathrm{g}}$).  The
argument of the logarithm is always positive, as is noted in
the inequality~(\ref{eq:logArgumentPositive1}).

\section{Equations of equilibrium}
\label{sec:equilibrium}

In this section, we use the calculus of variations to derive the
equations of equilibrium for a general ribbon model.  Thanks to our
parameterization based on directors, we do this simply by extending
the classical derivation of the equations of equilibrium for thin
elastic rods, using the principle of virtual work.  We have used a
similar approach in a recent paper~\cite[eqs.~9b and~A.2b]{Dias2013}
to derive the equations of equilibrium of a thin annular strip folded
along its central circle.

\subsection{Principle of virtual work for a ribbon}

If the external load is conservative, the equilibrium of the ribbon is
found by minimizing the total potential energy, which is the sum of
the potential energy associated with the external load and of the
elastic energy ${E}(\eta,\eta',\omega_{1})$ in
equation~\eqref{eq:Energy4}.  This is a constrained minimization
problem, and we need to extend the classical variational 
derivation~\cite{Cosserat1909,Cohen1966,%
Ericksen1970,%
Steigmann-Faulkner-Variational-theory-for-spatial-1993,%
Chouaieb-Kirchhoffs-problem-of-helical-solutions-of-uniform-2003,%
Audoly-Pomeau-Elasticity-and-geometry:-from-2010} of the equations of
equilibrium for thin rods to take into account the presence of the
\emph{kinematical constraints} $\mathcal{C}_{\mathrm{d}}$ and
$\mathcal{C}_{\mathrm{g}}$, and of an \emph{internal variable}
$\eta(S)$.

To handle the case of non-conservative loads, we derive the equations
of equilibrium using the more general framework of the principle of
virtual work.  A virtual motion of the ribbon is specified by a
virtual displacement $\hat{\mathbf{x}}(S)$ of the center-line, a
virtual rotation $\hat{\boldsymbol \psi}(S)$ of the orthonormal frame
of directors, a virtual variation $\hat{\mathbf{d}}_{i}(S)$ of the
directors, a virtual variation $\hat{\eta}(S)$ of the parameter
defining the direction of the generatrices, and the virtual changes of
material curvature and twisting strains $\hat{\omega}_{i}(S)$.

We start by noting that equations~(\ref{eq:adaptation}),
(\ref{eq:orthonormal}) and~(\ref{eq:Frame1-bis}) define what is known
as an inextensible Euler--Bernoulli rod.  When written in
incremental form, they yield relations between the virtual quantities
$\hat{\mathbf{x}}$, $\hat{\boldsymbol \psi}$, $\hat{\mathbf{d}}_{i}$
and $\hat{\omega}_{i}$.  In the principle of virtual work, these
relations are viewed as constraints, and are handled by a constraint
term $\mathcal{W}_{\mathrm{cEB}}$, which we call the virtual work of
the Euler--Bernoulli constraints.  Our treatment of these constraints
does not differ from the standard theory of inextensible
Euler--Bernoulli rods, and we shall therefore omit the details.

The virtual work of a ribbon includes the following contributions.  First,
the virtual internal work is equal to minus the first variation of the
elastic energy $\mathcal{W}_{\mathrm{i}} = -
\hat{{E}}$, as usual for elasticity problems.
Second, the external virtual work reads $\mathcal{W}_{e} = \int_{0}^{L}
(\mathbf{p}\cdot\hat{\mathbf{x}} +
\mathbf{c}\cdot\hat{\boldsymbol{\psi}})\,\textrm{d}S$ where
$\mathbf{p}$ and $\mathbf{c}$ are the density of applied force and
moment onto the center-line, per unit arc-length.  For static
problems, the virtual work of acceleration is zero.  The constraints
are treated by constraint terms involving Lagrange multipliers: the
constraints applicable to an inextensible Euler--Bernoulli model are
included in a constraint term $\mathcal{W}_{\mathrm{cEB}}$ as
explained above; the constraints $\mathcal{C}_{\mathrm{g}}$ and
$\mathcal{C}_{\mathrm{d}}$, which are specific to a ribbon, are
treated by two contributions $\mathcal{W}_{\mathrm{cg}} = \int_{0}^{L}
\lambda_{\mathrm{g}}\,\hat{\mathcal{C}}_{\mathrm{g}}\, \mathrm{d}S$
and $\mathcal{W}_{\mathrm{cd}} = \int_{0}^{L}
\lambda_{\mathrm{d}}\,\hat{\mathcal{C}}_{\mathrm{d}}\, \mathrm{d}S$,
which we call, respectively, the virtual work of the geodesic and developability
constraints.  Here, $\lambda_{\mathrm{g}}(S)$ and
$\lambda_{\mathrm{d}}(S)$ denote two Lagrange multipliers, and
$\hat{\mathcal{C}}_{\mathrm{i}}$ is the first variation of any of the
constraints, $i\in\{\mathrm{d},\mathrm{g}\}$.  Combining all the
contributions, we write the principle of virtual work as
\begin{multline}
    -\int_{0}^{L}\left(
    \frac{\partial \mathcal{E}}{\partial \eta'}\,\hat{\eta}'
    +\frac{\partial \mathcal{E}}{\partial \eta}\,\hat{\eta}
    +\frac{\partial \mathcal{E}}{\partial \omega_{1}}\,\hat{\omega}_{1}
    \right)\,\mathrm{d}S
    + \mathcal{W}_{\mathrm{e}}
    +\mathcal{W}_{\mathrm{cEB}}
    \cdots \\
    {}+\int_{0}^{L}\lambda_{\mathrm{g}}\,
    \frac{\partial \mathcal{C}_{\mathrm{g}}}{\partial \omega_{2}}\,
    \hat{\omega}_{2}\,\mathrm{d}S
    +\int_{0}^{L}\lambda_{\mathrm{d}}\,\left(
    \sum_{i=1}^3\frac{\partial \mathcal{C}_{\mathrm{d}}}{\partial \omega_{i}}\,
    \hat{\omega}_{i}
    +
    \frac{\partial \mathcal{C}_{\mathrm{d}}}{\partial \eta}\,
    \hat{\eta}
    \right)\,\mathrm{d}S
    = 0
    \textrm{,}
    \nonumber
\end{multline}
where $\mathcal{E}$ is the elastic energy density defined by $E=\int_{0}^{L}\mathcal{E}(\eta,\eta',\omega_{1})\,\mathrm{d}S$ .

Rearranging the terms, we have
\begin{equation}
    \mathcal{W}_{\eta}
    + \mathcal{W}_{\mathrm{i}}^\mathrm{rod}
    + \mathcal{W}_{\mathrm{e}}
    +\mathcal{W}_{\mathrm{cEB}}
    = 0
    \textrm{,}
    \label{eq:pvw}
\end{equation}
where we have grouped the terms depending on the internal variable
$\eta$,
\begin{equation}
    \mathcal{W}_{\eta}
    = 
    -\int_{0}^{L}\left[
    \frac{\partial \mathcal{E}}{\partial \eta'}
    \,\hat{\eta}'
    +
    \left(
    \frac{\partial \mathcal{E}}{\partial \eta}
    - \lambda_{\mathrm{d}}\,
    \frac{\partial \mathcal{C}_{\mathrm{d}}}{\partial \eta}
    \right)\,\hat{\eta}
    \right]\,\mathrm{d}S
    \textrm{,}
\end{equation}
and the terms depending on the virtual changes of 
strain,
\begin{equation}
    \mathcal{W}_{\mathrm{i}}^\mathrm{rod}
    =
    -
    \int_{0}^{L}
    \sum_{i=1}^3
    \left(
    \frac{\partial \mathcal{E}}{\partial \omega_{i}}
    -\sum_{j\in\{\mathrm{g},\mathrm{d}\}}
    \lambda_{j}\,\frac{\partial \mathcal{C}_{j}}{\partial \omega_{i}}
    \right)\,\hat{\omega}_{i}
    \;\mathrm{d}S
    \textrm{.}
    \label{eq:ivw-tmp1}
\end{equation}

Equation~(\ref{eq:pvw}) expresses the principle of virtual work for a
ribbon: equilibrium configurations are such that this equation is
satisfied for any kinematically admissible virtual displacement.
Making use of the strong similarity with the principle of virtual work
for a thin rod, we can now derive the equations of equilibrium for
ribbons easily.

\subsection{Equations of equilibrium}

The equation of equilibrium with respect to $\eta$ comes from the term
$\mathcal{W}_{\eta}$.  Integrating by parts and canceling the 
coefficient of $\hat{\eta}$, we find, as
in~\cite{Starostin2007},
    \begin{equation}
	\label{eq:etaEquil-tmp1}
	-\frac{\mathrm{d}}{\mathrm{d}S}\left(\frac{\partial\mathcal{E}}{\partial\eta'}\right)
	+\frac{\partial\mathcal{E}}{\partial\eta}
	-\lambda_{\mathrm{d}}\frac{\partial \mathcal{C}_{\mathrm{d}}}{\partial \eta}
	 = 0
	\textrm{.}
    \end{equation}
    Through its first term, this equation depends on the second
    arc-length derivative of $\eta$.  Boundary terms have been omitted: the
    derivation of the boundary condition for $\eta$ is irrelevant to
    the case of a closed ribbon, and is left to the reader.

We proceed to the second term $\mathcal{W}_{\mathrm{i}}^\mathrm{rod}$
in equation~(\ref{eq:pvw}).  We observe that it can be put into the usual form
of the internal virtual work of a thin
rod~\cite{Cohen1966,Steigmann-Faulkner-Variational-theory-for-spatial-1993,%
Chouaieb-Kirchhoffs-problem-of-helical-solutions-of-uniform-2003,%
Audoly-Pomeau-Elasticity-and-geometry:-from-2010}, namely
\begin{equation}
    \mathcal{W}_{\mathrm{i}}^\mathrm{rod} = -\int_{0}^{L} \mathbf{M}\cdot
    \sum_{i=1}^3(\hat{\omega}_{i}\,\mathbf{d}_{i})\,\mathrm{d}S
    \textrm{,}
    \label{eq:ivw-rod}
\end{equation}
when we identify the internal moment $\mathbf{M}(S)$ based on
equation~(\ref{eq:ivw-tmp1}):
\begin{multline}
    \label{eq:constit-tmp1}
    \mathbf{M} = \sum_{i=1}^3 \left(
    \frac{\partial \mathcal{E}}{\partial \omega_{i}}
    - \lambda_{\mathrm{d}}\,\frac{\partial 
    \mathcal{C}_{\mathrm{d}}}{\partial \omega_{i}}
    - \lambda_{\mathrm{g}}\,\frac{\partial 
    \mathcal{C}_{\mathrm{g}}}{\partial \omega_{i}}
    \right)\,\mathbf{d}_{i} \\
    =
     \left(
    \frac{\partial\mathcal{E}}{\partial\omega_1}
	-\eta\,\lambda_{\mathrm{d}}
    \right)\,\mathbf{d}_{1} + 
    \lambda_{\mathrm{g}}\,\mathbf{d}_{2}
    +
    \lambda_{\mathrm{d}}\,\mathbf{d}_{3}
    \textrm{.}
\end{multline}
This equation is one of the main results of our paper.  It expresses
the constitutive law of the ribbon in the language of thin rods.  The
case of a folded annular strip has been considered recently by the
same authors, and similar constitutive laws have been
obtained~\cite[eqs.~9b and~A.2b]{Dias2013}.  In
equation~(\ref{eq:constit-tmp1}), the first term in the parenthesis
yields the usual constitutive law $\mathbf{M} = \sum_{i}\frac{\partial
\mathcal{E}}{\partial \omega_{i}}\,\mathrm{d}_{i}$ for an unconstrained,
non-linearly elastic thin rod.  The two other terms are constraint
terms (the isotropic pressure term entering in the constitutive law of
an incompressible elastic solid is a constraint term of the same
kind).

Let us now consider the contributions to the principle of virtual work
that involve the virtual motion of the center-line and directors,
\emph{i.e.}\ those depending on $(\hat{\mathbf{x}}, \hat{\boldsymbol
\psi}, \hat{\mathbf{d}}_{i}, \hat{\omega}_{i})$.  These contributions
are $\mathcal{W}_{\mathrm{i}}^\mathrm{rod} + \mathcal{W}_{\mathrm{e}}
+\mathcal{W}_{\mathrm{cEB}}$, as the other contribution
$\mathcal{W}_{\eta}$ concerns the internal degree of freedom only.
These are exactly the same three terms as those entering in the
principle of virtual work for an inextensible Euler--Bernoulli rod.
It is well
known~\cite{Cohen1966,Steigmann-Faulkner-Variational-theory-for-spatial-1993,%
Chouaieb-Kirchhoffs-problem-of-helical-solutions-of-uniform-2003,%
Audoly-Pomeau-Elasticity-and-geometry:-from-2010} that the
corresponding equations of equilibrium are the Kirchhoff equations
expressing the balance of forces and moments on a small chunk of the
center-line,
\begin{highlightequation}
\begin{subequations}
    \label{eq:classicalKirchhoffEquilibrium}
    \begin{align}	
        \mathbf{R}'(S) + \mathbf{p}(S) &  = \mathbf{0}
	\label{eq:classicalKirchhoffEquilibrium-r}
	\textrm{,}
	\\
        \mathbf{M}'(S) + \mathbf{x}'(S)\times \mathbf{R}(S) + 
	\mathbf{c}(S) & =
	\mathbf{0} 
	\textrm{.}
	\label{eq:classicalKirchhoffEquilibrium-m}
    \end{align}
\end{subequations}
\end{highlightequation}
This remark saves us the effort of rederiving these equations of
equilibrium.  Note that there is no constitutive law associated with
the internal force $\mathbf{R}(S)$, as $\mathbf{R}(S)$ is the
Lagrange multiplier associated with the Euler--Bernoulli constraint in
equation~(\ref{eq:adaptation}).

To sum up, an elastic ribbon is governed by the same equations as an
inextensible Euler--Bernoulli rod, up to two small changes: (\emph{i})
the presence of the internal degree of freedom $\eta$ yields the new
equation of equilibrium~(\ref{eq:etaEquil-tmp1}), and (\emph{ii}) the
presence of constraint terms in the constitutive law; see
equation~(\ref{eq:constit-tmp1}).


\subsection{Complete set of equations for an elastic ribbon}

The complete set of equations for the equilibrium of a ribbon are
summarized as follows.  The geometry and the natural shape of the
ribbon are prescribed by the geodesic curvature
$\kappa_{\mathrm{g}}(S)$, by the edge functions $V_{\pm}(\eta,S)$, and
by the tensor $\mathbf{Q}_{\mathrm{r}}$ defined in
equation~(\ref{eq:MomentQ}).  The external loading is prescribed by
the functions $\mathbf{p}(S)$ and $\mathbf{c}(S)$, subject to the
global balance of forces and moments.  The equilibrium
configuration of the ribbon is sought in terms of the following
\emph{unknowns}: the center-line $\mathbf{x}(S)$, the directors
$\mathbf{d}_{i}(S)$, the parameter $\eta(S)$ capturing the direction
of the generatrices, and the Lagrange multipliers $M_{2}(S) =
\lambda_{\mathrm{g}}(S)$ and $M_{3}(S) = \lambda_{\mathrm{d}}(S)$.

The \emph{geometrical} equations are the compatibility of the tangent
and the third director in equation~(\ref{eq:adaptation}), the
orthonormality of the directors in equation~(\ref{eq:orthonormal}),
and the developability constraints in
equations~(\ref{eq:geodesicConstraint}--\ref{eq:developabilityConstraint}).
The rotation gradient $\omega_{i}(S)$ is found using
equation~(\ref{eq:Frame1-bis}).

In terms of the unknowns, one expresses the coordinate
$V_{\mathrm{c}}$ of the striction curve using
equation~(\ref{eq:Vcritical}),
and the curvature tensor $\mathbf{K}_{0}$ along the center-line using
equation~(\ref{eq:Curvature2FormMaterial0}).
Based on the definition of the energy functional in
equation~(\ref{eq:Energy4}), one can then calculate the internal
moment by the \emph{constitutive law}, which we rewrite from 
equation~(\ref{eq:constit-tmp1}) as
\begin{highlightequation}
\begin{equation}
    \mathbf{M}(S) = 
    \left(
    \frac{\partial\mathcal{E}}{\partial\omega_1}
	-\eta\,M_{3}
    \right)\,\mathbf{d}_{1}(S) + 
    M_{2}(S)\,\mathbf{d}_{2}(S)
    +
    M_{3}(S)\,\mathbf{d}_{3}(S)
    \textrm{.}
    \label{eq:constit-final}
\end{equation}
\end{highlightequation}
The \emph{equations of equilibrium} are the standard Kirchhoff 
equations~(\ref{eq:classicalKirchhoffEquilibrium})
together with equation~(\ref{eq:etaEquil-tmp1}) for the equilibrium of 
$\eta$, which we rewrite as
\begin{highlightequation}
\begin{equation}
    	\label{eq:equilibrium-repeat}
	-\frac{\mathrm{d}}{\mathrm{d}S}\left(\frac{\partial\mathcal{E}}{\partial\eta'}\right)
	+\frac{\partial\mathcal{E}}{\partial\eta}
	-M_{3}\,\omega_{1} = 0
	\textrm{.}
\end{equation}
\end{highlightequation}

If the ribbon is closed, these equations must be complemented by
periodic boundary conditions\footnote{Finding the conditions for a
continuous curve in space to be closed is not a trivial problem and it
may not have a close solution.  This problem was posed by N. V. Efimov
\cite{Efimov1947} and W. Frenchel \cite{Frenchel1951}.  Frenchel posed
the problem asking what are the necessary and sufficient conditions of
closure given the curvature and torsion of a space curve.  By
integrating the Frenet--Serret equations, the result yields an
infinite series of integrals with no closed form
\cite{Cheng-Chung1981}.}.  If it is open, boundary conditions
corresponding to the type of support must be enforced --- their
derivation has been left to the reader.

The equations tagged by a star, such as
equations~(\ref{eq:constit-final}) and~(\ref{eq:equilibrium-repeat})
above, form the complete set of equations governing the equilibrium of
a ribbon.

\section{Special cases}
\label{sec:specialcases}

We review ribbon models known in the literature, that are specific to
the case $\kappa_{\mathrm{g}} = 0$, in the light of our general model.

\subsection{Naturally straight, rectangular ribbons}

This model was first introduced by Wunderlich in his study of the
shape of a M\"obius band~\cite{Wunderlich1962}.  The M\"obius band is
a naturally flat and straight ribbon having a constant width $w$: we set $\mathbf{Q}_{\mathrm{r}} = \mathbf{0}$, 
$\kappa_{\mathrm{g}} = 0$. Then, by equation~(\ref{eq:Vcritical}), 
the coordinate of the striction curve is $V_{\mathrm{c}} = -1/\eta'$. 
The relevant edge functions are given by equations~(\ref{eq:rectangular-ribbon})
as $V_{\pm} = \pm w/2$. Inserting into the energy functional in 
equation~(\ref{eq:Energy4}), we have
\begin{equation}
    \label{eq:Energy5}
    {E}_{\mathrm{W}}
    =\frac{D\,w}{2}\int_0^L{\omega_1}^2\,\left(1+\eta^2\right)^2\,\frac{1}{\eta'\,w}\,
    \ln\left(\frac{1+\eta'\,w/2}{1-\eta'\,w/2}\right)\;\mathrm{d}S,
\end{equation}
where we have used
$\tr\left({\mathbf{K}_{0}}^2\right)={\omega_1}^2\left(1+\eta^2\right)^2$.
Wunderlich's energy functional is recovered~\cite{Wunderlich1962}. It
has been analyzed recently and is relevant to the elastic M\"obius
strip~\cite{MahadevanKeller-The-shape-of-a-Mobius-band-1993,%
Starostin2007}, and to open developable ribbons~\cite{Korte2010}.

Starostin and van der Heijden ~\cite{Starostin2007} have worked out
the corresponding equations of equilibrium (for the naturally flat,
rectangular ribbon).  Here, we have recovered their results as a
special case: their equations~[4] are identical\footnote{The
quantities $(2w,\mathbf{t},\mathbf{n},\mathbf{b},\tau,\kappa, \eta,g,
\mathbf{M}, \mathbf{F}, M_{t},M_{b})$ in their notation must be
identified with the quantities $(w,\mathbf{d}_{3}, -\mathbf{d}_{2},
\mathbf{d}_{1}, \omega_{3}, \omega_{1},\eta, {E}, -\mathbf{M},
-\mathbf{R}, -M_{3}, -M_{1})$ in our notation, respectively.  The
last four minus signs introduced here arise because they use
a non-standard convention for the sign of the internal force
$\mathbf{R}$ and moment $\mathbf{M}$ --- by contrast, we use the
usual convention that $\mathbf{R}$ and $\mathbf{M}$ measure the force
and moment applied across an imaginary cut by the downstream part of
the ribbon onto the upstream part, where `downstream' and
`upstream' refer to the direction of increasing arc-length coordinate $S$.} to the Kirchhoff
equations~(\ref{eq:classicalKirchhoffEquilibrium}) derived above;
their equations~[5] are the constitutive law~(\ref{eq:constit-final})
and the equilibrium for the internal variable $\eta$.

\subsection{Sadowsky's limit: narrow rectangular ribbons}

The limiting case of Wunderlich's energy in
equation~\eqref{eq:Energy5} corresponding to $w\to 0$ was derived by
Sadowsky~\cite{Sadowsky30} much before Wunderlich.  It reads
\begin{equation}
    \label{eq:Energy6}
    {E}_{S} =\frac{D\,w}{2}\int_0^L{\omega_1}^2
    \left(1+\eta^2\right)^2\mathrm{d}S
    \textrm{.}
\end{equation}
This \emph{narrow strip} model applies to a rectangular, naturally
flat ribbon, when the deformation is sufficiently small for the
striction curve to remain far from the physical edge of the ribbon,
$|V_{\mathrm{c}}|\gg w$.  This model captures the geometrically
non-linear coupling between bending and twisting modes.
It has been applied to the statistical mechanics of developable
ribbons~\cite{Giomi2010} and to the analysis of elastic strips
comprising a central fold~\cite{Dias2013}.

The equations governing the equilibrium of Wunderlich's strip model,
derived in reference~\cite{Starostin2007}, apply as a special case to
the equilibrium of Sadowsky's narrow strip model.  An alternative
derivation appears in reference~\cite[section 2]{Dias2013}.

\subsection{Helical ribbons}
\label{ssec:helicalRibbons}

The case of a naturally helical ribbon
has been studied recently~\cite{Starostin2008,%
Chopin-Kudrolli-Helicoids-Wrinkles-and-Loops-2013}.
Its geodesic curvature is zero, $\kappa_{\mathrm{g}} = 0$.  Assuming
that the helical shape is stress-free, the
natural curvature tensor reads, from
equation~(\ref{eq:Curvature2FormMaterial0}),
\begin{equation}
    \mathbf{K}_{\mathrm{r}} =
    -\omega_{1}^{\mathrm{r}}\,\mathbf{d}_3
    \otimes\mathbf{d}_3
    +\omega_{3}^{\mathrm{r}}\,
    (
    \mathbf{d}_3\otimes\mathbf{d}_1
    +
    \mathbf{d}_1\otimes\mathbf{d}_3
    )
    -\frac{({\omega_{3}^{\mathrm{r}})^2}}{\omega_{1}^{\mathrm{r}}}\,\mathbf{d}_1\otimes\mathbf{d}_1
    \nonumber
\end{equation}
where $\omega_{1}^\mathrm{r}$ and $\omega_{3}^\mathrm{r}$ are the two
parameters defining the radius and the step of the helix.  The
equations of equilibrium, derived
in~\cite{Starostin2008}
by the same method as in~\cite{Starostin2007}, can be recovered from
the general model derived above.

When subjected to moderate forces and moments compatible with the
helical symmetries, these ribbons remain helical --- such helical
configurations are relevant to the opening of chiral seed
pods~\cite{Armon2011a}, a phenomenon driven by residual internal
stresses~\cite{Wu2013}.  In the presence of larger or less symmetric
loads, non-helical solutions are
possible~\cite{Starostin2008,%
Chopin-Kudrolli-Helicoids-Wrinkles-and-Loops-2013}.

\section{Illustrations}
\label{sec:illustrations}

With the aim to illustrate our results and motivate further studies,
we present a few problems that could be solved using the equations
derived in this work.

\subsection{Buckling of a cylindrical ribbon}
\label{ssec:slabCylindrical}

\begin{figure}
    \centering
    \includegraphics[width=2.5in]{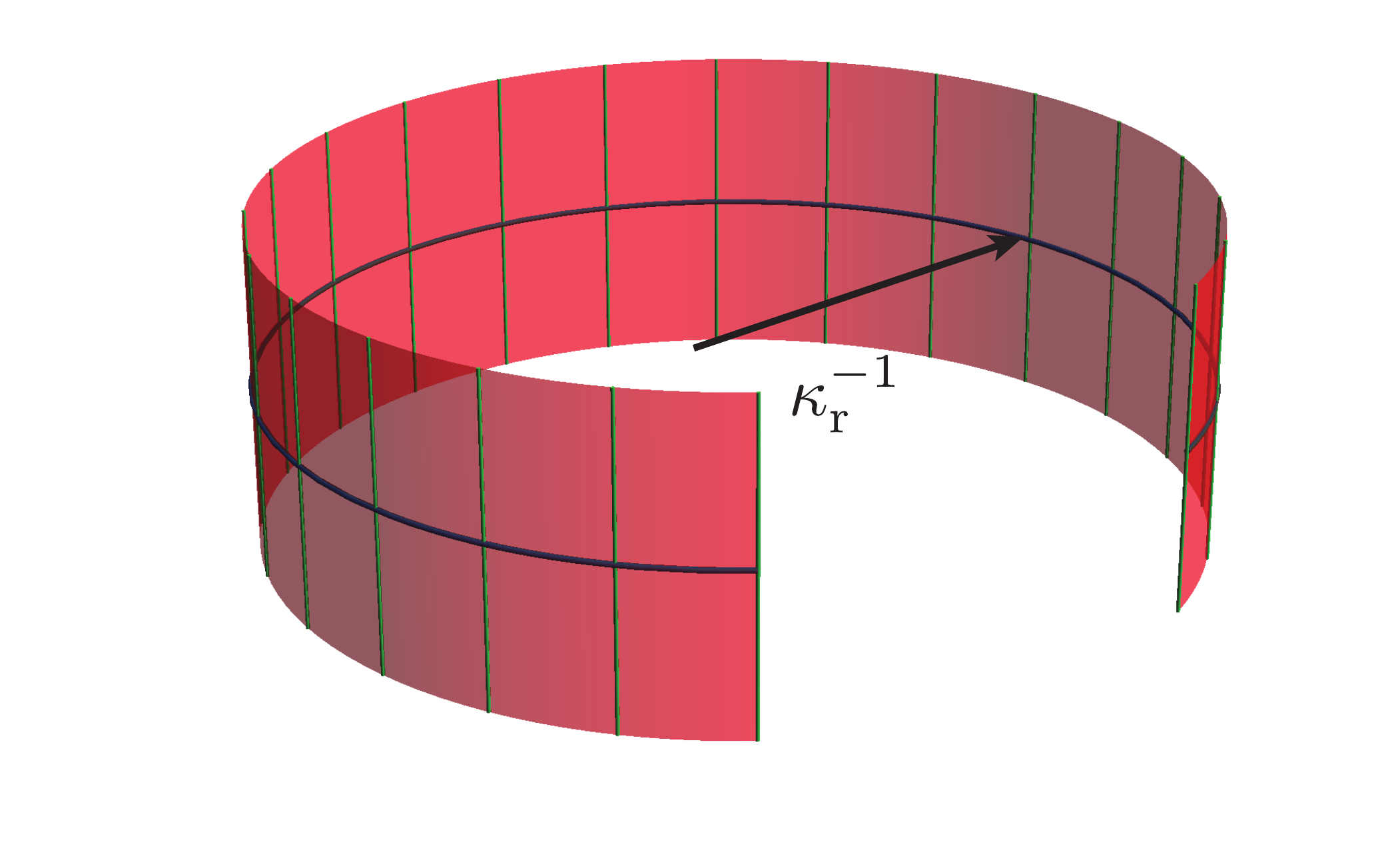}
    \caption{\label{fig:Fig4} A ribbon cut out in a cylindrical shell
    having natural curvature $\kappa_\mathrm{r}$
    (\S\ref{ssec:slabCylindrical}).}
\end{figure}

We consider the cylindrical ribbon shown in figure~\ref{fig:Fig4}.
Its natural curvature is denoted by $\kappa_{\mathrm{r}}$.  When laid
flat, the ribbon is a rectangle of size $L\times w$.  In its natural
configuration, the direction supported by its width $w$ is aligned
with the axis of the cylinder: its natural curvature is an
out-of-plane curvature $\kappa_{\mathrm{r}} = \omega_{1}^\mathrm{r}$,
its geodesic curvature being zero, $\kappa_{\mathrm{g}} = 0$.
The tensor of natural curvature reads
$\mathbf{K}_\mathrm{r}=-\kappa_\mathrm{r}\,\mathbf{d}_3 \otimes
\mathbf{d}_3$, as shown by inserting $\omega_{1} =
\kappa_{\mathrm{r}}$ and $\eta = \omega_{3} / \omega_{1} = 0$ into
equation~(\ref{eq:Curvature2FormMaterial0}).  Using the definition of
the internal moment $\mathbf{Q}_{\mathrm{r}}$ in
equation~\eqref{eq:MomentQ}, we have
\begin{equation}
    \label{eq:MomentQEx}
    \mathbf{Q}_\mathrm{r}=-\kappa_\mathrm{r}\,\left(\mathbf{d}_3
    \otimes\mathbf{d}_3+\nu\,\mathbf{d}_1\otimes\mathbf{d}_1\right).
\end{equation}
By equation~\eqref{eq:Energy4}, the energy functional governing this 
strip model reads
\begin{equation}
    \label{eq:Energy7}
    {E} =\frac{D\,w}{2}\int_0^L
    \left[{\omega_1}^2\left(1+\eta^2\right)^2\,\frac{1}{w\,\eta'}\ln\left(\frac{1+\eta'w/2}{1-\eta'w/2}\right)
    -2\,\kappa_\mathrm{r}\,\omega_1\left(1+\nu\,\eta^2\right)\right]\mathrm{d}S
    ,
\end{equation}
which is simply Wunderlich's energy, complemented by a term depending
on natural curvature.  For a narrow strip (Sadowsky's limit,
$\eta'w\rightarrow0$), this becomes
\begin{equation}
    \label{eq:Energy8}
    \begin{split}
	{E} & 
	=\frac{D\,w}{2}\,\int_0^L\left[{\omega_1}^2\left(1+\eta^2\right)^2
	-2\,\kappa_\mathrm{r}\,\omega_1\left(1+\nu\,\eta^2\right)\right]
	\;\mathrm{d}S
	\\
	& 
	=\frac{D\,w}{2}\,\int_0^L
	\left[
	\left({\omega_1}\left(1+\eta^2\right)
	- \kappa_\mathrm{r}\right)^2
	- {\kappa_{\mathrm{r}}}^2
	\right]
	\;\mathrm{d}S
	\textrm{.}
    \end{split}
\end{equation}
	
The buckling of a closed ribbon of this type, caused by a mismatch of
the natural curvature and the curvilinear length of the ribbon
($\kappa_{\mathrm{r}} \neq 2\pi/L$), is studied in another
article~\cite{SeffenAudoly-2014} based on the equations derived here.

\subsection{Overcurved and undercurved annular ribbons}
\label{ssec:overcurvedAnnular}

\begin{figure}[tbp]
    \centerline{\includegraphics[width=4.5in]{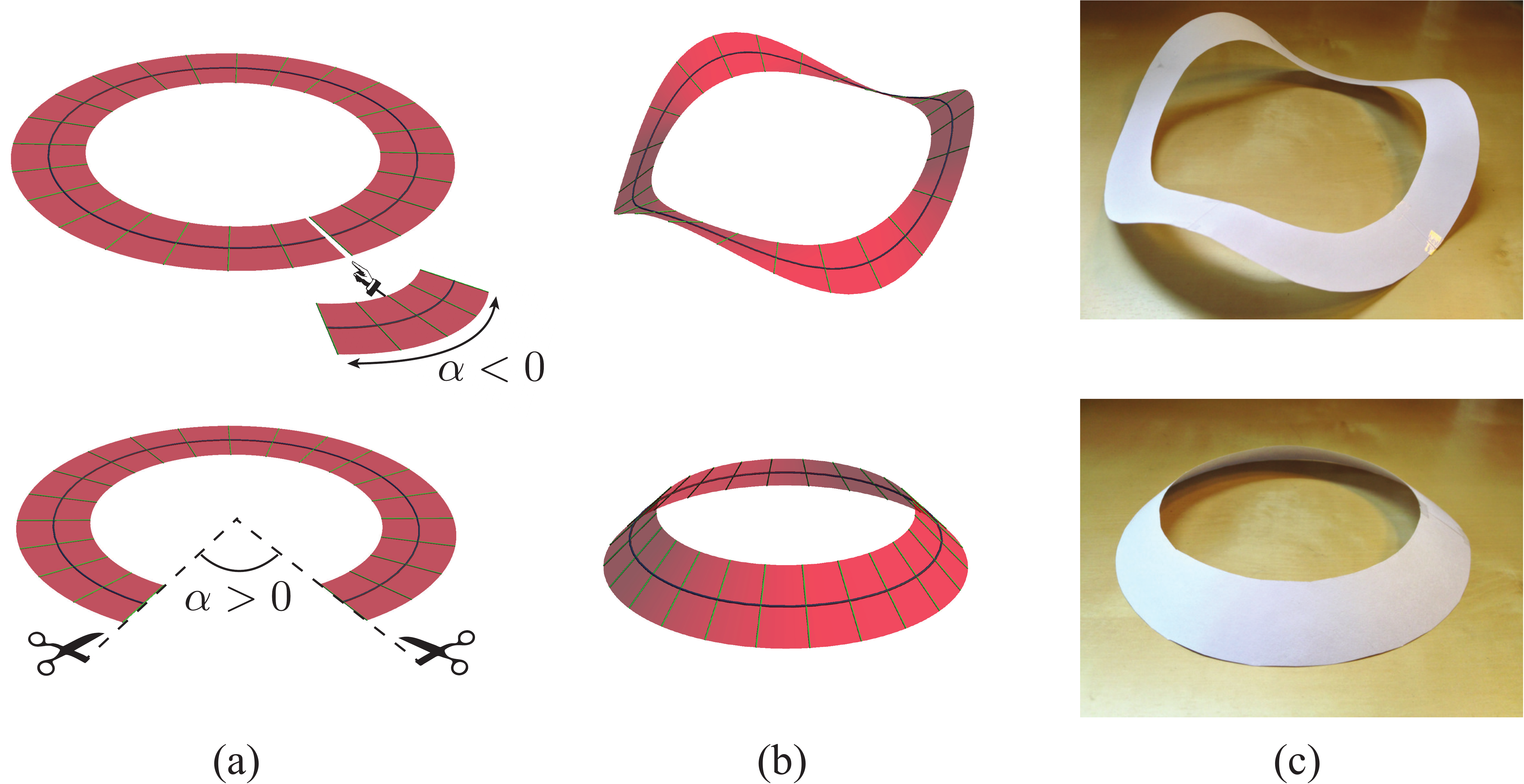}}
    \caption{Undercurved (top row) and overcurved (bottom row) annuli
    (\S\ref{ssec:overcurvedAnnular})~: (a) preparation of the two
    states, (b) sketch of typical equilibrium shapes, (c) experiments
    using paper models.
}
    \protect\label{fig:Fig5}
\end{figure}

The undercurved and overcurved annuli shown in figure~\ref{fig:Fig5} are simple
examples of ribbons having non-zero natural curvature.
They can be obtained as a paper model by cutting out an annulus from a
sheet of paper.  Assume the reference line to be middle line of the annular strip: 
its curvature defines the geodesic curvature
$\kappa_{\mathrm{g}}\neq 0$.  If a sector with angle $\alpha$ is
removed from the annulus and the two newly formed ends are pasted
together, the annulus becomes overcurved: its curvilinear length $L =
(2\,\pi - \alpha)/\kappa_{\mathrm{g}}$ is less than that of a circle
with curvature $\kappa_{\mathrm{g}}$, namely
$2\pi/\kappa_{\mathrm{g}}$.  The undercurved annuli can be achieved by
pasting together two identical annuli, each one being such that
$\alpha < \pi$ --- the effective value of $\alpha$ after pasting is
then negative, $\alpha<0$.  In the presence of undercurvature or
overcurvature, the annulus will buckle out-of-plane.  This problem is
a variant of the problem of the \emph{folded} annular strip, which we
have studied recently~\cite{Dias2012a,Dias2013}.  The buckling of the
annular ribbon without a fold has not yet been analyzed to the best of
our knowledge.  This can be done using the equations derived in this
article.

\subsection{3D kirigami from 2D cut-out patterns}
\label{ssec:cutouts}

\begin{figure}[tbp]
    \centerline{\includegraphics[width=4.5in]{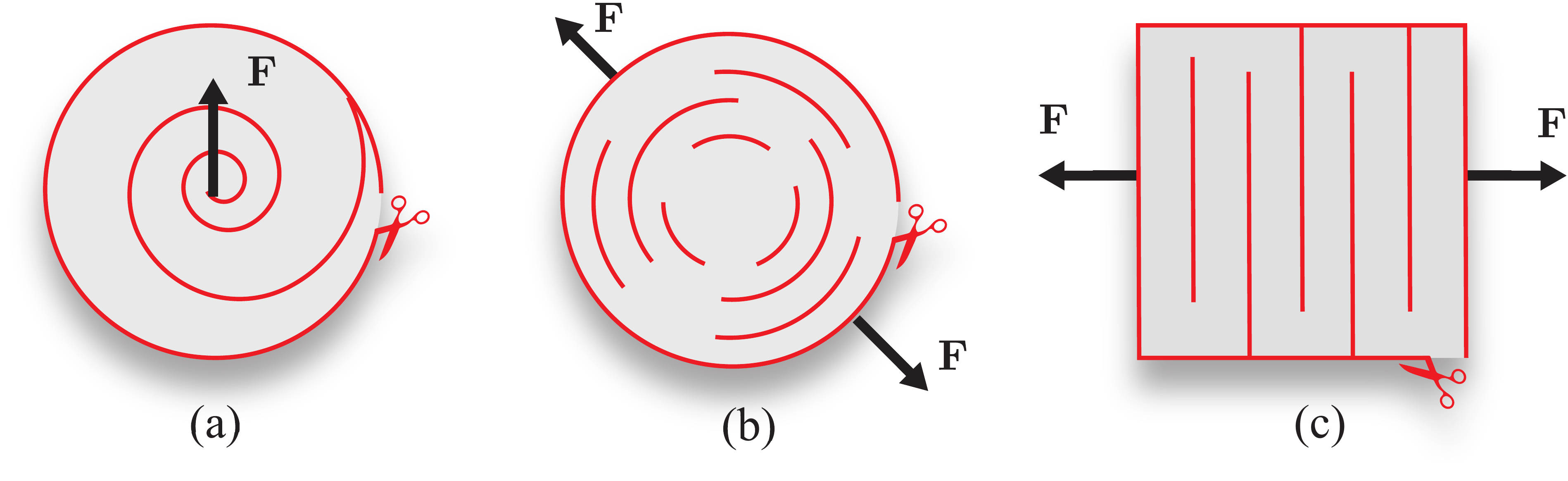}}
    \caption{Patterns cut out in a piece of paper can produce various
    3D shapes when pulled (\S\ref{ssec:cutouts}): (a) spiral cut
    pattern, (b) concentric, circular cut patterns, (c) straight,
    alternating cut patterns. 
}
    \protect\label{fig:Fig9}
\end{figure}

Complex 3D shapes can be obtained by pulling on a thin sheet of paper
that has been cut along arbitrary lines; see figure~\ref{fig:Fig9}.
This special form of kirigami (the art of \emph{cutting paper})
involves tuning the geometry of the cuts to produce various 3D shapes.
For instance, the alternated concentric cuts shown in figure~\ref{fig:Fig9}b are the basis of a commercial paper model that
produces a bowl-like shape.
In all these examples, the center-line is not a geodesic.  The example
shown in figure~\ref{fig:Fig9}c, for instance, is made by patching
rectangular strips together, and it can be viewed as a single strip
having a variable width and a zigzagging center-line.  The unfolding of
these 3D kirigami could be analyzed using the general ribbon model
derived in this article.

\section{Conclusion}
\label{sec:conclusion}

We have presented a general theory of ribbons that fits into the
well-established framework of thin elastic rods.  By working with a
frame of directors (material frame) instead of the theory of
Frenet--Serret frame, we could extend the energy functional and the
equilibrium equations to a general ribbon geometry.  In particular,
geodesic curvature, out-of-plane curvature, and a variable width have
been taken into account.  Instead of using the inextensibility
conditions to eliminate degrees of freedom, as in previous work, we
have treated them as constraints in the sense of the calculus of
variations.  This allowed us to view an elastic ribbon as a special
kind of a thin elastic rod --- namely, a kinematically constrained,
hyperelastic rod possessing an internal degree of freedom.  We could
also explain the deep similarities between the theories of ribbons and
thin rods.

This unifying view bridges the gap between classical rod theory and
the theory of elastic ribbons --- the latter has been developed
largely as an independent subject so far.  This makes it possible to
reuse the large body of numerical and analytical methods available for
thin rods.  Numerical methods for simulating elastic ribbons, for
instance, are currently limited to the calculation of non-linear
equilibria, using numerical continuation, see
e.g.~\cite{Starostin2007}.  Numerical continuation is a powerful tool
but its use can be quite impractical.  Even though this is standard
for thin
rods~\cite{ColemaSwigon-Theory-of-Supercoiled-Elastic-Rings-2000}, we
are not aware of any simulation method that can predict the dynamics
of ribbons, or account for self-contact.  When extended to the
dynamical case by including the virtual power of acceleration, the
constrained variational formulation presented here offers a natural
way of porting existing numerical methods for the dynamics of thin
rods, such as the finite-element
method~\cite{YangTobiasOlson-Finite-element-analysis-of-DNA-supercoiling-1993}
or methods based on discrete differential
geometry~\cite{Bergou-Wardetzky-EtAl-Discrete-Elastic-Rods-2008}, to
ribbons.

Here we have focused on developable configurations.  Under large loads
involving a combination of tension and twisting, ribbons can adopt
non-developable configurations~\cite{Green-The-Elastic-Stability-of-a-Thin-II-1937,%
Mockensturm-The-Elastic-Stability-of-Twisted-2001,%
Chopin-Kudrolli-Helicoids-Wrinkles-and-Loops-2013}.  In future work,
it would be interesting to extend the parameterization of ribbons
presented here to account for deviations from the constraint of local area preservation.

We would like to thank E.~Fried for his detailed comments and
suggestions on the manuscript.  MAD thankfully acknowledges support
from NSF Grant No.\ CBET-0854108.

\appendix

\section{Curvature tensor of a developable surface}
\label{SIA}

Here, we use the notation introduced in section~\ref{ssec:devtransf}
to prove the geometrical identities announced in
sections~\ref{ssec:constraints} to~\ref{ssec:curvatureTensor},
relevant to developable surfaces.  We consider a general ruled
surface, enforce the condition of developability, and derive the
expression of the curvature tensor at an arbitrary point on the
surface.  By doing so, we extend the expressions obtained by
Wunderlich~\cite{Wunderlich1962} and by Starostin and van der Heijden~\cite{Starostin2007}
to account for the geodesic curvature $\kappa_{\mathrm{g}}$ of the
center-line.

Let us first calculate the tangent vectors at an arbitrary point 
$\mathbf{y}(S,V)$ of the surface:
\begin{subequations}
    \label{eq:tangentVectors}
    \begin{align}
        \mathbf{y}_{,S}(S,V) &  = \mathbf{d}_{3}(S) + V\,\mathbf{q}'(S)
        \label{eq:tangentVectors-S}\\
        \mathbf{y}_{,V}(S,V) & = \mathbf{q}(S)
	\textrm{,}
	\label{eq:tangentVectors-V}
    \end{align}
\end{subequations}
where $\mathbf{q}'(S)$ denotes the total derivative of $\mathbf{q}$ 
defined in equation~(\ref{eq:EmbedRibbon2}) as $\mathbf{q} = 
\eta\,\mathbf{d}_{3} + \mathbf{d}_{1}$.

Using equation~(\ref{eq:Frame1}), we write the following equation
\begin{equation}
    \mathbf{q}' = \eta\,\omega_{2}\,\mathbf{d}_{1}
    + (\omega_{3} - \eta\,\omega_{1})\,\mathbf{d}_{2}
    + (\eta'-\omega_{2})\,\mathbf{d}_{3}
    \textrm{,}
\end{equation}
which implies
\begin{equation}
    \mathbf{q} \times \mathbf{q}' = (\eta\,\omega_{1}- \omega_{3}) 
    \,\mathbf{q}^\perp + \frac{1}{V_{\mathrm{c}}}\,\mathbf{d}_{2}
    \label{eq:qCrossqPrime}
\end{equation}
where $V_{\mathrm{c}}$ is the quantity defined by 
equation~(\ref{eq:Vcritical}), and $\mathbf{q}^\perp$ is the vector
\begin{equation}
    \mathbf{q}^\perp = \mathbf{d}_{2} \times \mathbf{q}  = 
    -\mathbf{d}_{3} + \eta\,\mathbf{d}_{1}
    \textrm{.}
\end{equation}
Later on, we shall show that $\mathbf{q}^\perp$ is a vector
perpendicular to $\mathbf{q}$ lying in the plane tangent to the
surface; hence the notation.

The classical condition for a ruled surface to be
developable~\cite[section
3.II]{Spivak-A-comprehensive-introduction-vol3-1999} is that the
following three vectors are linearly dependent: the tangent
$\mathbf{d}_{3}$ to the center-line (called the directrix in the
context of the geometry of surfaces), the vector $\mathbf{q}$ spanning
the generatrices, and its derivative with respect to the arc-length
along the center-line.  This is expressed by $( \mathbf{q}\times
\mathbf{q}')\cdot \mathbf{d}_{3} = 0$.  In view of
equation~(\ref{eq:qCrossqPrime}), this yields $\eta\,\omega_{1} =
\omega_{3}$, which is the constraint of developability announced in
equation~(\ref{eq:developabilityConstraint-cancels}).

As a result, $\mathbf{q}' \cdot \mathbf{d}_{2} = 0$, and so
$\mathbf{y}_{,S}\cdot \mathbf{d}_{2} = 0$.  On the other hand,
equation~(\ref{eq:tangentVectors-V}) shows that $\mathbf{y}_{,V}\cdot
\mathbf{d}_{2} = 0$.  The director $\mathbf{d}_{2}(S)$ is orthogonal
to both tangents: $\mathbf{d}_{2}$ is a unit normal at any point of
the developable surface.

The element of area on the surface reads
\begin{equation}
    \mathrm{d} a = \left|\mathbf{y}_{,S}\times 
    \mathbf{y}_{,V}\right|\,\mathrm{d}S\,\mathrm{d}V
    = \left| \mathbf{d}_{3}\times \mathbf{q} + V \mathbf{q}'\times 
    \mathbf{q}\right| \,\mathrm{d}S\,\mathrm{d}V
    = \left|
    \left(1-\frac{V}{V_{\mathrm{c}}}\right)\,\mathbf{d}_{2}
    \right|\,\mathrm{d}S\,\mathrm{d}V
\end{equation}
Noting that $1-V/V_{\mathrm{c}} > 0$ by 
the inequality~(\ref{eq:logArgumentPositive2}), we arrive at the result 
announced in equation~(\ref{eq:Element}).

To compute the curvature tensor $\mathbf{K}(S,V)$, we note that the 
direction of the generatrix is a principal direction of zero 
curvature, since the surface is developable. Therefore, there exists 
some scalar field $k(S,V)$ such that
\begin{equation}
    \mathbf{K} = k\,\mathbf{q}^\perp \otimes \mathbf{q}^\perp
    \textrm{.}
    \label{eq:Kqperp}
\end{equation}
The quantity $k$ in equation above can be found by contracting with
$\mathbf{y}_{,S}$ on both sides of the equation to give:
\begin{equation}
    \mathbf{y}_{,S}\cdot \mathbf{K}\cdot \mathbf{y}_{,S}
    =k\,\left(\mathbf{q}^\perp \cdot 
    \left(\mathbf{d}_{3}+V\,\mathbf{q}'\right)\right)^2
    = k\,\left(-1+\frac{V}{V_{\mathrm{c}}}\right)^2
    \textrm{.}
    \label{eq:k-v1}
\end{equation}

By the definition of the curvature tensor (second fundamental
form)~\cite{Spivak-A-comprehensive-introduction-vol3-1999},
the left-hand side of the resulting identity is the normal projection of the second derivative 
$\mathbf{y}_{,SS}$:
\begin{equation}
    \mathbf{y}_{,S}\cdot \mathbf{K}\cdot \mathbf{y}_{,S}
    =
    \mathbf{y}_{,SS}\cdot \mathbf{d}_{2} = \left(\mathbf{d}_{3}' + 
    V\,\mathbf{q}''\right)\cdot \mathbf{d}_{2} = -\omega_{1}
    +V\,\left(
    \frac{\mathrm{d}(  \mathbf{q}'\cdot \mathbf{d}_{2})}{\mathrm{d}S}
    - \mathbf{q'}\cdot \mathbf{d}_{2}'
    \right)
    \textrm{.}
\end{equation}
In this equation, $\mathbf{q}'\cdot \mathbf{d}_{2} = \omega_{3}- 
\eta\,\omega_{1}= 0$ by the developability condition, and 
$\mathbf{q}'\cdot \mathbf{d}_{2}' = \mathbf{q}'\cdot 
(\boldsymbol{\omega}\times \mathbf{d}_{2}) = \mathbf{q}'\cdot  
(\omega_{1}\, \mathbf{q}\times \mathbf{d}_{2}) = - 
\omega_{1}\,\mathbf{d}_{2}\cdot(\mathbf{q}\times \mathbf{q}') = 
-\omega_{1} / V_{\mathrm{c}}$. Therefore,
\begin{equation}
    \mathbf{y}_{,S}\cdot \mathbf{K}\cdot \mathbf{y}_{,S}
    = - \omega_{1} \,\left(1-\frac{V}{V_{\mathrm{c}}}\right)
    \label{eq:k-v2}
\end{equation}

From equations~(\ref{eq:k-v1}) and~(\ref{eq:k-v2}), we can solve for $k$, giving
$k = -\omega_{1} /(1-V/V_{\mathrm{c}})$. Inserting this result into
equation~(\ref{eq:Kqperp}) yields the expression of the
curvature tensor announced in
equations~(\ref{eq:Curvature2FormMaterial})
and~(\ref{eq:Curvature2FormMaterial0}).  The curvature tensor keeps
the same form as in the case of zero geodesic
curvature~\cite{Wunderlich1962,Starostin2007} provided that the proper
definition of $V_{\mathrm{c}}$ in terms of $\kappa_{\mathrm{g}}$ is
used; see equation~(\ref{eq:Vcritical}).

\bibliographystyle{spmpsci}      



\end{document}